\documentclass[12pt, english]{article} 

\usepackage{geometry}
\usepackage{setspace}
\usepackage{caption}
\AtBeginCaption{\doublespacing}
\usepackage{lineno}
\usepackage{multirow} 
\renewcommand{\tabcolsep}{0.12cm} 

\usepackage{amsmath} 

\usepackage{array} 
\usepackage{MnSymbol} 
\usepackage{graphicx} 
\usepackage{placeins} 

\usepackage{natbib} 
\usepackage{authblk} 
\usepackage{threeparttable} 

\usepackage[compact]{titlesec} 
\titleformat{\section}{\normalfont\bfseries}{\thesection}{1em}{}
\titleformat{\subsection}{\normalfont\bfseries}{\thesubsection}{1em}{}
\titleformat{\subsubsection}{\normalfont\bfseries}{\thesubsubsection}{1em}{}
\titlespacing*{\section}{0pt}{*0}{-6pt}
\titlespacing*{\subsection}{0pt}{*0}{-6pt}
\titlespacing*{\subsubsection}{0pt}{*0}{-6pt}

\begin{document}
\begin{spacing}{1.9} 

\title{Differentiating the L\'evy walk from a composite correlated random walk}
\date{}
\author[1,\footnote{Email: marie.auger-methe$@$ualberta.ca}]{Marie Auger-M\'eth\'e}
\author[1,\footnote{Email: derocher$@$ualberta.ca}]{Andrew E. Derocher}
\author[2,\footnote{Email: michael.plank$@$canterbury.ac.nz}]{Michael J. Plank}
\author[3,\footnote{Email: ecodling$@$essex.ac.uk}]{Edward A. Codling}
\author[1,4,\footnote{Email: mark.lewis$@$ualberta.ca}]{Mark A. Lewis}

\affil[1]{\small{Department of Biological Sciences, University of Alberta, Edmonton, Canada}} \affil[2]{Department of Mathematics and Statistics, University of Canterbury, Christchurch, New Zealand}
\affil[3]{Department of Mathematical Sciences, University of Essex, Colchester, United Kingdom}
\affil[4]{Centre for Mathematical Biology, Department of Mathematical and Statistical Sciences, University of Alberta, Edmonton, Canada}
\maketitle

\noindent \textbf{Word count:}
6718

\noindent \textbf{Short running head:} Differentiating random search models

\noindent \textbf{List of online material:} Appendixes A B \& C

\noindent \textbf{Manuscript type:} Standard article

\newpage
\begin{flushleft} 
\begin{abstract}

1. Understanding how to find targets with very limited information is a topic of interest in many disciplines. In ecology, such research has often focused on the development of two movement models: i) the L\'evy walk and; ii) the composite correlated random walk and its associated area-restricted search behaviour. Although the processes underlying these models differ, they can produce similar movement patterns. Due to this similarity and because of their disparate formulation, current methods cannot reliably differentiate between these two models.

2. Here, we present a method that differentiates between the two models. It consists of likelihood functions, including one for a hidden Markov model, and associated statistical measures that assess the relative support for and absolute fit of each model.

3. Using a simulation study, we show that our method can differentiate between the two search models over a range of parameter values. Using the movement data of two polar bears (\textit{Ursus maritimus}), we show that the method can be applied to complex, real-world movement paths.

4. By providing the means to differentiate between the two most prominent search models in the literature, and a framework that could be extended to include other models, we facilitate further research into the strategies animals use to find resources.

\end{abstract}


\section{Introduction}

Search strategies that allow targets to be found with very limited information are relevant to diverse fields of study \citep{RefWorks:949}. In particular, they have received much attention in the animal movement literature, where the two most prominent random search models are the L\'evy walk and the composite correlated random walk (CCRW), with its associated area-restricted search behavior \citep{RefWorks:347,RefWorks:929,RefWorks:997}. The L\'evy walk is a popular but controversial movement model that is defined as a random walk with a power-law distribution describing the step length frequency \citep{RefWorks:440, RefWorks:430,RefWorks:942,RefWorks:943,Pyke2014}. This distribution has a characteristic heavy tail that allows for arbitrarily long step lengths. L\'evy walks are sometimes inaccurately referred to as L\'evy flights in the movement literature \citep[see][]{Pyke2014}. Area-restricted search (also known as area-concentrated search) is the process whereby animals restrict their movement to the vicinity of recent captures, and is particularly useful in heterogeneous environments \citep{RefWorks:392,RefWorks:962}. Area-restricted search is one of two behaviors often modeled with CCRWs or similar composite random walks \citep{RefWorks:962,RefWorks:440}. Such two-behavior models typically consist of `extensive' and `intensive' phases, and are often used to identify foraging events or locate food patches from movement data \citep[e.g.,][]{RefWorks:389,RefWorks:997,RefWorks:986}. Each behavior is related to a specific part of the landscape. The intensive search behavior is triggered by the encounter of a food item. This behavior is called area-restricted search because the animal uses low speed and large turning angles to remain within a food patch and thus increase the probability of detecting prey. The extensive search behavior is resumed after repeated unsuccessful searches. It uses fast and nearly straight movement to find the next food patch. Both the L\'evy walk and CCRWs with area-restricted search have been claimed to be optimal under certain conditions \cite[][but see \citealt{RefWorks:941}]{RefWorks:962,RefWorks:434} and both have empirical support \citep[e.g.,][]{RefWorks:997,RefWorks:942}.

Although the processes underlying these two search models differ widely in their biological interpretation, their movement patterns are similar and difficult to differentiate. Many have argued that the CCRW could be confounded with the L\'evy walk \citep{RefWorks:440,RefWorks:432,RefWorks:438,RefWorks:928} and the disparate formulation of these models hinders their direct comparison \citep{RefWorks:930}. In response, new methods to identify the L\'evy walk have been developed
\cite[][but see \citealt{RefWorks:1004}]{RefWorks:963,RefWorks:964,RefWorks:990}. However, these improved methods cannot be used to quantify the evidence for the CCRW. Quantifying the level of evidence for each model is important as it both reduces the potential for misidentification and allows for a more comprehensive analysis of search strategies. Recently, methods have been proposed that simultaneously fit the L\'evy walk and models approximating the CCRW \citep{RefWorks:936, RefWorks:932}. Although these methods represent significant improvements over previous approaches, they do not fully represent the CCRW as they lack turning angles and temporal correlation in behaviors. Turning angles are an essential part of movement and are crucial for distinguishing between the two movement behaviors found in the CCRW \citep{RefWorks:962,Pyke2014}. Temporal correlation in behaviors is an inherent characteristic of the CCRW because it is required to create the tortuous movement that allows the animal to remain in a food patch.

Here, we present a new method for differentiating between the movement patterns of the L\'evy walk and CCRW. In the proposed method, the CCRW is represented by a hidden Markov model (HMM) that incorporates turning angles and behavioral persistence \citep[similar to][]{RefWorks:991}. For comparability, the L\'evy walk and two null models are modified to incorporate turning angles. Likelihood functions for these models are created because they are essential for a set of statistical measures that assess both the relative and absolute support for each model. Using a simulation study, we show that our method can be used to successfully differentiate between the movement patterns of L\'evy walks and CCRWs and to assess the relative and absolute fit of the models. We demonstrate the applicability of our method by applying it to the movement paths of two polar bears (\textit{Ursus maritimus}).

\section{Methods}

\subsection{Development of the proposed method}
\label{sec:OurMethod}

Our proposed method consists of likelihood functions representing each search model and statistical measures that use these likelihoods to assess the support for each model.

\subsubsection{Likelihood functions}
\label{sec:Models}

Our likelihood functions use the information from both movement measures of a step: $(l_i, \theta_i)$. The step length, $l_i$, is defined as the distance between the starting and ending locations of the $i^{th}$ step. The turning angle, $\theta_i$, is defined as the angle of a step relative to the previous step direction. Only steps with a sufficient number of locations to measure both a step length (i.e., requires two locations) and a turning angle (i.e., requires three locations) are included. In addition, because we focus on the case where animals are moving and potentially searching (i.e., not performing behaviours such as resting) and because L\'evy walks only model step lengths greater than 0 (see below), we exclude steps with identical start and end points. Excluding steps is possible because the models either assume that each measure of movement is independent and identically distributed or, in the case of the HMM, are built to handle missing steps. In this section, we present the development of the likelihood functions representing a CCRW, L\'evy walk, and two null models. The four likelihoods differ mainly in the probability density functions (PDFs) chosen to describe the step length and turning angle frequencies.

A CCRW is a combination of two random walks, representing two behavioral modes. Similar to \citet{RefWorks:438}, we describe the tortuous movement of the intensive search (hereafter denoted with subscript $\textsc{i}$) with a Brownian walk (BW) and the directed movement of the extensive search (hereafter denoted with subscript $\textsc{e}$) with a correlated random walk (CRW). The BW and CRW are two common models that differ in their turning angle distribution. While an animal following a BW has no preferred turning direction, one following a CRW has a tendency to continue in the same direction as the previous step \citep{RefWorks:925}.

For each behavior, we define the turning angle frequency with one of two circular PDFs. To represent the intensive search as a BW, we use a circular uniform distribution, $v_0(\theta)$ (Appendix A: Table A.1).  For the extensive search, we chose the von Mises distribution. This distribution was used in recent studies comparing L\'evy walks and CCRWs \citep{RefWorks:438,RefWorks:932}. The von Mises distribution has two parameters: $\alpha$, which is the location parameter and can be interpreted as the mean angle between steps; and $\kappa$, which is the scale parameter and can be interpreted as the size of the directional correlation. To represent the extensive search as a CRW, we set $\alpha_{\textsc{e}}=0$ and estimate $\kappa_{\textsc{e}}$. This von Mises distribution is similar to a circular version of the Gaussian distribution centered at 0 \citep{RefWorks:926} and is represented as $v(\theta|\kappa_{\textsc{e}})$ (Appendix A: Table A.1).

For each behavior, we model the step lengths with a slightly modified exponential distribution, $\phi(l|\lambda, a)$ (Appendix A: Table A.1). This modified exponential distribution is often used as an alternative to the L\'evy walk \citep[e.g.,][]{RefWorks:928, RefWorks:924,RefWorks:963} and was used in previous attempts to compare multiphasic movement to the L\'evy walk \citep{RefWorks:936}. The exponential distribution defines the probability of a step length as exponentially decreasing with increasing size. The modified exponential distribution starts at the minimum step length, $a$, rather than starting at 0. This modification is equivalent to applying the exponential distribution to the difference between the step length and the minimum step length, $l-a$. The distribution often used to model L\'evy walks requires the minimum step length, $a$, to be greater than 0 \citep{RefWorks:924,RefWorks:926,RefWorks:940}. As such, the data sets used in L\'evy walk studies exclude step length of 0 (i.e., when the animal remains stationary). The modified exponential distribution can model data sets that exclude step lengths of 0 and thus makes our CCRW directly comparable to L\'evy walk models. 

Each exponential distribution has two parameters to estimate: the minimum step length, $a$, and the rate parameter, $\lambda$. While the minimum step length, $a$, is assumed to be the same for both behaviors, $\lambda$ differs between behaviors: $\lambda_{\textsc{i}}$ and $\lambda_{\textsc{e}}$. We can interpret $\lambda$ as the inverse of the mean step length \citep{RefWorks:926}, or more precisely as the inverse of the mean difference between step lengths and the minimum step length, $\lambda = n/\sum_{i=1}^{n}(l_i-a)$. Thus a difference between $\lambda_{\textsc{i}}$ and $\lambda_{\textsc{e}}$ captures differences in the distances moved in each behavior. By combining the step length and turning angle distributions, we get the following observation PDFs associated with each behavior:
\begin{linenomath*}
\begin{equation}
p_{\textsc{i}}(l_i,\theta_i)= \phi(l_i| \lambda_{\textsc{i}},a) \; v_{0}(\theta_i),
\label{eqn:obsProbX}
\end{equation}
\end{linenomath*}
and
\begin{linenomath*}
\begin{equation}
p_{\textsc{e}}(l_i,\theta_i)= \phi(l_i| \lambda_{\textsc{e}},a) \; v(\theta_i|\kappa_{\textsc{e}}).
\label{eqn:obsProbY}
\end{equation}
\end{linenomath*}

The observation PDFs describing the movement of each behavior are combined through what is referred as a mixing distribution. The choice of mixing distribution is an important difference between our model and the previous attempts to compare multiphasic movement to the L\'evy walk \citep{RefWorks:936,RefWorks:932}. Previous models combined the observation probabilities through an independent mixing distribution, where the probability of intensively searching and that of extensively searching are independent of previous probabilities and constant through time. Although these models provide good approximations to the movement of an animal that has two behaviors, they do not represent the temporal correlation in behaviors that a HMM can provide. Behavioral persistence is crucial when modeling the CCRW without including environmental variables as the trigger for behavioral switches. In our case, we implicitly represent the spatial correlation that a patchy landscape would create with first-order temporal correlation in behavior. Thus, unlike models with an independent mixing distribution, the order of the observations is important in a HMM.

We used the methods of \citet{RefWorks:933} to create a HMM from our observation probabilities. The mixing distribution is a first-order Markovian process, which models the transition between the behavior of consecutive steps through the transition probability matrix:
\begin{linenomath*}
\begin{equation}
\setstretch{1}
\boldsymbol{\Gamma} =
\begin{pmatrix}
 \gamma_{\textsc{ii}} & 1 - \gamma_{\textsc{ii}}  \\
 1-\gamma_{\textsc{ee}} & \gamma_{\textsc{ee}}  \\
\end{pmatrix},
\label{eqn:tpm}
\end{equation}
\end{linenomath*}
where $\gamma_{\textsc{ii}}$ and $\gamma_{\textsc{ee}}$ are the probabilities of remaining in the intensive and extensive search behaviors, respectively, and $1-\gamma_{\textsc{ii}}$ and $1-\gamma_{\textsc{ee}}$ are the probabilities of switching from intensive to extensive and from extensive to intensive, respectively. Because the duration of each movement phase follows a geometric distribution, $1/( 1-\gamma_{\textsc{ii}})$ and $1/( 1-\gamma_{\textsc{ee}})$ can be interpreted as the mean number of steps the animal remains in the intensive and extensive search, respectively. Thus, an animal that remains on average more than two steps in the same search behavior will have $\gamma_{\textsc{ii}}$ and $\gamma_{\textsc{ee}} > 0.5$. As the probability of being in a behavior depends on the previous probabilities, we need to define the initial probability of being in each behavior:
\begin{linenomath*}
\begin{equation}
\setstretch{1}
\boldsymbol{\delta} =
\begin{pmatrix}
 \delta_{\textsc{i}} & 1-\delta_{\textsc{i}} \\
\end{pmatrix},
\label{eqn:initDist}
\end{equation}
\end{linenomath*}
where $\delta_{\textsc{i}}$ and $1-\delta_{\textsc{i}}$ are the probabilities of starting in the intensive and extensive search behaviors, respectively. The likelihood of the CCRW is:
\begin{linenomath*}
\begin{equation}
L_{\textsc{ccrw}}(\boldsymbol{\Theta} | \boldsymbol{l},\boldsymbol{\theta}) = \boldsymbol{\delta} \boldsymbol{P}(l_1,\theta_1) \; \prod^{n}_{i=2} ( \boldsymbol{\Gamma} \boldsymbol{P}(l_i,\theta_i) ) \; \boldsymbol{1},
\label{eqn:LikelihoodHMM}
\end{equation}
\end{linenomath*}
where $\boldsymbol{1}$ is a column vector of ones and $\boldsymbol{P}(l_i,\theta_i)$ is the observation probability matrix that incorporates the probability of being in each behavior as defined by \ref{eqn:obsProbX} and \ref{eqn:obsProbY}:
\begin{linenomath*}
\begin{equation}
\setstretch{1}
\boldsymbol{P}(l_i,\theta_i) =
\begin{pmatrix}
 p_{\textsc{i}}(l_i,\theta_i) & 0 \\
 0 & p_{\textsc{e}}(l_i,\theta_i)  \\
\end{pmatrix}.
\label{eqn:obsProbMatrix}
\end{equation}
\end{linenomath*}
The expanded formula of the likelihood can be found in Table \ref{tab:Models}. As mentioned above, we used the von Mises and exponential distributions in our HMM because they have been used in related research \citep{RefWorks:438,RefWorks:936,RefWorks:932}. However, this approach can be generalized. The HMM framework is flexible, and other turning angle and step length distributions can be used to create CCRWs \citep[e.g., wrapped Cauchy and Weibull distributions, see:][]{RefWorks:991}.

To make the likelihood of the L\'evy walk comparable to the CCRW, we used a PDF for the turning angle in addition to the PDF that is generally used to describe the step lengths of the L\'evy walk (Table \ref{tab:Models}). Following others \citep[e.g.,][]{RefWorks:480,RefWorks:932,Pyke2014}, we assume that the turning angle of the L\'evy walk is uniform. Thus, we used the same circular uniform PDF, $v_{0}(\theta)$, as described above (Appendix A: Table A.1). Two step length PDFs can be used to describe the L\'evy walk. One represents the pure L\'evy walk, the other represents the truncated L\'evy walk (TLW). Unlike the pure L\'evy walk, the TLW places an upper bound on the size of possible step lengths, making it biologically plausible \citep{RefWorks:929}. As a result, the TLW is often used as the L\'evy walk model for animal movement \citep[e.g.,][]{RefWorks:943}. The step length PDF of the TLW is the truncated Pareto, $\psi_{\textsc{t}}(l|\mu_{\textsc{t}},a,b)$ (Appendix A: Table A.1). This distribution has three parameters to estimate: the shape parameter, $\mu_{\textsc{t}}$, which increases the probability of long step length as it decreases, the minimum step length, $a$, and the maximum step length, $b$, which represents either the greatest step length an animal can make or the greatest distance between prey encounters. The likelihood for the TLW is:
\begin{linenomath*}
\begin{equation}
L_{\textsc{tlw}}(\boldsymbol{\Theta} | \boldsymbol{l},\boldsymbol{\theta}) = \prod^{n}_{i=1} \; \psi_{\textsc{t}}(l_i|\mu_{\textsc{t}},a,b)  \; v_{0}(\theta_i)
\end{equation}
\end{linenomath*}
While we focused on the TLW in the main text, we present analyses of the pure L\'evy walk in Appendix A.

To verify that the complexity associated with the CCRW and TLW is required to explain the data, it is important to compare these models against simpler ones. Therefore we used likelihood functions for two simpler models: the BW and CRW (Table \ref{tab:Models}). The BW is a null model representing an individual moving randomly in space, while the CRW represents movement with directional persistence. These models are closely related to the null models used in L\'evy walk studies \citep{RefWorks:480,RefWorks:430} and use the same circular and exponential PDFs as the observation PDFs of the CCRW (\ref{eqn:obsProbX} and \ref{eqn:obsProbY}). The truncated version of the exponential distribution is sometimes used as a null model in L\'evy walk studies \citep[e.g.,][]{RefWorks:430} and we present analyses of the truncated version of these two models in Appendix A.

\subsubsection{Statistical measures}
\label{sec:fitModel}

To assess the support for each search model, we used the likelihood functions described above with a set of statistical measures. First, we computed the maximum likelihood estimates (MLEs) of the model parameters and calculated their confidence intervals through likelihood surface analyses. Second, we compared the fit of the models with Akaike Information Criterion ($AIC$) and Akaike weights. Finally, we tested the absolute fit of the models through analyses of pseudo-residuals. We performed these analyses with R 3.1.1 \citep{Rcite}. The R code and Rcpp source code for the R package we have developed is available on GitHub (https://github.com/MarieAugerMethe/CCRWvsLW).

We used maximum likelihood to estimate the parameters of the models described above (Table \ref{tab:AllPar}). We used known analytical solutions for the MLE of $a$ and $b$ \citep{RefWorks:940}. For the remaining parameters, we used numerical optimizing functions and, in the case of our CCRW, we used the Expectation-Maximization (EM) algorithm described by \citet{RefWorks:933}. We used the EM algorithm for our CCRW because it could be readily coded with Rcpp \citep{Eddelbuettel2011}. The resulting Rcpp algorithm was orders of magnitude faster than using R's numerical optimizers to directly maximize the likelihood. Given that we fit our CCRW to 77 700 simulations, computational efficiency was an important consideration (see next section). A disadvantage of using the EM algorithm over the direct numerical maximization is the need to estimate $\delta_{\textsc{i}}$ \citep{RefWorks:933}, a parameter with little biological relevance. While both methods generally produce similar results, the EM algorithm is harder to code than the numerical maximization of the likelihood \citep{MacDonald2014}. Thus, while our fast Rcpp EM algorithm was required for our simulation study, the direct numerical maximization of the CCRW likelihood would have been easier to implement and would be an adequate solution to fit a CCRW to empirical data. Newly developed HMMs may be more easily implemented using direct numerical maximization of the likelihood \citep[e.g.,][]{RefWorks:991}.

To estimate the confidence intervals of the parameters, we used the quadratic approximation described by \citet{RefWorks:934}. This method uses the Hessian of the negative log likelihood at its minimum value. As the analytical solution of $a$ and $b$ is to use the minimum and maximum observed step lengths \citep{RefWorks:940} and the estimated value from the EM algorithm for $\delta_{\textsc{i}}$ depends only on the observations of the first step \citep{RefWorks:933}, it is difficult to estimate confidence intervals for these three parameters. We only provide point estimates for them.

The main goal of our likelihood functions is to identify which model fits the data best. To do so, we compared the relative fit of the models using $AIC_c$ and Akaike weights \citep{RefWorks:231}. The model with the lowest $AIC_c$ is considered to be the best model. To measure the weight of evidence the best model has over the other models, we calculated Akaike weights, $w$, from the $AIC_c$ values of the models \citep{RefWorks:231}. Akaike weight values vary between 0 and 1, with a weight close to 1 suggesting that the data strongly support the model over the other models investigated.

As the best model according to $AIC_c$ and Akaike weights can still be a poor representation of the data, it is important to verify its absolute fit \citep{RefWorks:930}. In the context of L\'evy walk analyses, the suggested test of absolute fit is a G-test \citep{RefWorks:430,RefWorks:924}, a test that assumes that observations are independent of one another. This assumption is violated in the case of the CCRW because this model incorporates temporal autocorrelation. Hence, we modified the test of absolute fit by applying the G-test to pseudo-residuals rather than to observations. We used ordinary uniform pseudo-residuals, which are residuals that account for the interdependence of observations and are uniformly distributed when the model adequately describes the data \citep{RefWorks:933}. We performed a G-test that compares the observed frequency of these pseudo-residuals to a discretized uniform distribution. To reduce the potential bias associated with bins that have small expected values, we used William's correction and ensured that each bin had 10 expected pseudo-residuals \citep{RefWorks:225}. We applied the G-test to the pseudo-residuals of step length and turning angle independently and subsequently combined their p-values using Fisher's method \citep{RefWorks:225}. One can further investigate the absolute fit of the models by looking for the presence of autocorrelation in the pseudo-residuals. Appendix B describes pseudo-residuals and the test of absolute fit in more detail.

\subsection{Simulation study}
\label{sec:Sim}

We used simulations to assess whether our method can differentiate between the TLW and CCRW (code also available on Github: https://github.com/MarieAugerMethe/CCRWvsLW). Because parameter values affect the resemblance of these models \citep{RefWorks:1004}, we simulated the CCRW and TLW on a range of parameter values. For each set of parameters, we ran 50 simulations. Each simulation created a movement path of 500 biologically relevant steps (i.e., representing animal movement decisions, for which a constant time interval is not assumed, see Appendix C for simulations investigating alternative conditions). For each simulation, we used our proposed method to estimate the parameter values and calculate the Akaike weights of all models. This allowed us to verify that the method could accurately estimate parameters and appropriately differentiate between models. To assess whether the true model was rejected at the appropriate $\alpha$-level, we also calculated the p-value of the absolute fit test associated with the simulated model.

To simulate the CCRW, we initialized the movement path by selecting the starting behavior, either $I_1$ or $E_1$, using a Bernoulli distribution with probability of being in the intensive search behavior defined by $\delta_{\textsc{i}}$. If the behavior was the intensive search, we randomly selected a turning angle from a circular uniform distribution and a step length from an exponential distribution with $\lambda_{\textsc{i}}$. If the behavior was the extensive search, we randomly selected a turning angle from a von Mises distribution with $\kappa_{\textsc{e}}$ and a step length from an exponential distribution with $\lambda_{\textsc{e}}$. After selecting the turning angle and step length for the first step, we selected the next behavioral state with a Bernoulli distribution that used the transition probability appropriate for the current behavior (i.e., $\gamma_{\textsc{ii}}$ if in intensive search and $\gamma_{\textsc{ee}}$ if in extensive search).  As for the first step, we then selected a step length, a turning angle and the behavioral state for the next step from the appropriate distributions. This process was continued until the last step of the movement path. Our CCRW has seven parameters (Tables \ref{tab:Models} and \ref{tab:AllPar}). We fixed the values of $\delta_{\textsc{i}}$, $\lambda_{\textsc{i}}$, and $a$, to 0.5, 0.01, and 1, respectively. We varied the value of $\kappa_{\textsc{e}}$ to $(0.5, 1, 5, 10)$, that of $\lambda_{\textsc{e}}$ to $(0.01, 0.005, 0.001, 0.0005, 0.0001)$, that of $\gamma_{\textsc{ii}}$ to $(0.6, 0.7, 0.8, 0.9)$, and that of $\gamma_{\textsc{ee}}$ to $(0.1, 0.2,..., 0.9)$. By choosing $\lambda_{\textsc{i}} \geq \lambda_{\textsc{e}}$, the step lengths from the extensive search behavior were either the same length or longer on average than those from the intensive search. We chose the values of $\gamma_{\textsc{ii}}$ to be $>0.5$ because the intensive search of the CCRW is efficient only if the animal remains multiple steps in a food patch. In contrast, we allowed $\gamma_{\textsc{ee}}$ to be $<0.5$ because an efficient extensive search for a food patch can be produced in one step. All 720 combinations of these parameters were simulated.

For each step of the TLW simulations, we randomly selected a turning angle from a circular uniform distribution, and a step length from a truncated Pareto distribution. The TLW has three parameters (Tables \ref{tab:Models} and \ref{tab:AllPar}). We set $a=1$ and varied the value of $\mu_{\textsc{t}}$ to (1.1, 1.2, ..., 2.9) and $b$ to (100, 1000, 10000). All 57 combinations of these parameters were simulated.

\subsection{Application to empirical data}
\label{sec:methodEmp}

To demonstrate its usefulness, we applied our method to the movement path of two polar bears from the Western Hudson Bay, Manitoba, Canada (data available on the University of Alberta Education \& Research Archive: http://hdl.handle.net/10402/era.40993). These two adult females were captured in September 2010 using the standard immobilization techniques \citep[][approved by the University of Alberta BioSciences Animal Policy and Welfare Committee - Protocol \#6001004]{RefWorks:427} and were collared with Gen IV collars from Telonics (Telonics Inc., Mesa, AZ, U.S.A). The collars were programmed to collect GPS locations at varying frequencies throughout the year. We used data from April 2011, the longest period with high frequency locations (location taken every 30 minutes) and a period where bears search for food \citep{RefWorks:1010,RefWorks:1067}. These two bears were on the sea ice during this period.

We applied our method to the data from each individual separately after estimating biologically relevant steps from the raw GPS data. Multiple techniques can be used to transform locations collected at regular time intervals into a time-series of biologically relevant steps \citep[e.g.,][]{RefWorks:996,RefWorks:928,RefWorks:1006}. In part for its ease of use, we used the local turn technique, which creates one step out of all consecutive sampled steps with a turning angle smaller than a threshold angle \citep[see][]{RefWorks:928}. We have shown elsewhere that using these types of techniques can results in misidentifying CCRWs for the L\'evy walk \citep{RefWorks:928,RefWorks:932}. However, such misidentification occurs mainly when high threshold angles are used \citep{RefWorks:928,RefWorks:932}. We chose a threshold angle of $10^{\circ}$, meaning that any sampled step within the $20^{\circ}$ forward sector is interpreted as part of a biologically relevant step. Thus resulting steps are created from movement in the same general direction and the threshold is small enough that it is unlikely to result in misidentification. We applied our method to empirical data to demonstrate how to interpret results and to show the performance of our method with real animal movement paths, which, unlike simulated movement, are complicated by factors such as missing data. See Appendix C for a simulation study exploring the effects of the local turn method on movement paths and a description of how missing data are handled.

\section{Results}

\subsection{Simulation results}
\label{sec:SimRes}

The Akaike weights could differentiate the L\'evy walk from a CCRW. When the CCRW was simulated, $81.6\%$ of the Akaike weight values of the CCRW exceeded 0.99 and the Akaike weight values of TLW never exceeded 0.01 (Fig. \ref{fig:aw_4M}A). Although the CCRW simulations were never misidentified as a TLW, $11.9\%$ of the summed Akaike weight values of the null models, $w_{\textsc{BW}} + w_{\textsc{CRW}}$, exceeded 0.5. This only occurred when the step length distribution of the extensive search was relatively close to that of the intensive search, $\lambda_{\textsc{e}} = (0.01, 0.005)$. In addition, this was generally limited to cases when the tendency to continue in the same direction was relatively low, $\kappa_{\textsc{e}} \leq 1$. When the TLW was simulated, 96.7\% of the Akaike weight values of the TLW exceeded 0.99 (Fig. \ref{fig:aw_4M}B). While 3.3\% of the Akaike weight value of the CCRW exceeded 0.01, only 0.1\% exceeded 0.5. Note that, due to underflow, we were unable to estimate the $AIC_c$ value of the CCRW for 0.3\% of the simulations. The Akaike weights results presented above and MLE results below ignore all problematic simulations.

In addition to differentiating between the two models, our method was capable of recovering the parameter values of the CCRW and TLW. As some parameter estimates can help identify whether the data are consistent with the L\'evy walk or with a CCRW with an efficient area-restricted search behaviour, it is important for our method to adequately estimate their values. The CCRW requires specific values for $\gamma_{\textsc{ii}}$ and $\kappa_{\textsc{e}}$ to be an efficient search model. The values of $\lambda_{\textsc{i}}$ and $\lambda_{\textsc{e}}$ can help further characterize the CCRW used by the animal. The TLW requires specific values for $\mu_{\textsc{t}}$ to be an efficient L\'evy walk. For most parameters of the simulated CCRW and TLW, the median of the estimated values was close to their true value (Figs. \ref{fig:HMM_MLE} and \ref{fig:TLW_MLE}). There were three exceptions. First, the estimated values of the initial probability of being in the intensive search of the CCRW, $\delta_{\textsc{i}}$, approached either 0 or 1, not 0.5 (Fig. \ref{fig:HMM_MLE}F). Second, some estimates of the minimum step length, $a$, were positively biased, and those of maximum step length, $b$, were negatively biased (Figs. \ref{fig:HMM_MLE}G and \ref{fig:TLW_MLE}B-C). Third, similar to the Akaike weights, the estimates of most parameters of the CCRW were less accurate when the movement patterns of the two behaviors were similar. Specifically, the estimates were less reliable when the simulations values of $\lambda_{\textsc{e}}$ were relatively close to $\lambda_{\textsc{i}}$. The estimated values of most parameters were much closer to the true value when simulations with $\lambda_{\textsc{e}} = (0.01, 0.005)$ were excluded. While the point estimates were generally reliable, the $95\%$ confidence intervals, as estimated with the quadratic approximation, had a tendency to be too narrow and excluded the correct simulation value more that $5\%$ of the time (Appendix C Table C.1) and often overlapped with the parameter space boundary (CCRW: $9.2 \%$, TLW: $5.6\%$), we thus used the more computationally intensive profile likelihood for the empirical results.

Finally, our tests of absolute fit had rejection rates adequate for the selected $\alpha$-level of 0.05 (p-value $< 0.05$). The proportion of simulated CCRWs that were rejected from being CCRW was 0.062. Similarly, the proportion of simulated TLWs that were rejected from being TLW was 0.067.

Appendix C shows that our method is affected by the local turn method, a technique used to transform raw GPS data into biologically relevant steps. Because the local turn method amalgamates all consecutive sampled steps with less than a threshold angle, and thus removes small turning angles, parameter estimates were heavily affected. In particular, $\kappa_{\textsc{e}}$ estimates were negatively biased. Using the local turn method also affected the test of absolute fit. When movement paths are transformed with such method, as in the case of our polar bear data, the test of absolute fit should only use the pseudo-residuals associated with the step lengths. The results in Appendix C demonstrate that the local turn method strongly affects the parameter estimates and the test of absolute fit. However, when used with a small threshold angle, this technique did not  decrease appreciably the capacity of our method to distinguish between the TLW and a CCRW.

\subsection{Empirical results}
\label{sec:EmpRes}

The best model for the two empirical movement paths was our CCRW (Table \ref{tab:EmpModelFitH}). For Bear 2, the Akaike weights indicated that the CCRW was a much better model than the other alternatives. However, the Akaike weight of the CCRW for Bear 1 was only 0.55, with some evidence that the CRW may have been a more parsimonious description of the movement data (Table \ref{tab:EmpModelFitH} and Fig. \ref{fig:2Bears}). While the best model was the CCRW, both movement paths were significantly different from it (Table \ref{tab:EmpModelFitH}). The movement path of Bear 1 was also significantly different from the CRW ($p < 0.01$). A visual representation of the fit of the models is presented in Fig. \ref{fig:2Bears}.

To identify whether the movement paths were consistent with the best model, we investigated the parameter estimates of the CCRW. For Bear 2, all parameters were consistent: $\gamma_{\textsc{ii}} > 0.5$ and $\kappa_{\textsc{e}} > 0$ (Table \ref{tab:AllPar}). In contrast, not all parameters for Bear 1 were consistent with the CCRW. While $\kappa_{\textsc{e}} > 0$ as expected, $\gamma_{\textsc{ii}} < 0.5$.

\section{Discussion}

\label{sec:Disc}

Through the analysis of TLW and CCRW simulations, we have demonstrated that our method can differentiate between the movement patterns of a L\'evy walk and CCRW. The Akaike weights identified the correct search model, except for a few instances. The Akaike weights also distinguished the TLW and CCRW from our two null models. The rare exceptions occurred when both the intensive and extensive search behaviors of the CCRW simulations had similar step length and turning angle distributions. This was expected. Other methods developed to distinguish the intensive from the extensive search are also less efficient when the movement of these behaviors are similar \citep{RefWorks:986}. When the two behaviors are similar, models describing them as one behavior can be sufficient. The ability of our method to differentiate between a CCRW and null models would likely increase with sample size.

The simulation analyses also indicated that most parameter estimates of the TLW and CCRW were reliable. The estimates of the important parameters of both models (e.i., $\gamma_{\textsc{ii}}$, $\lambda_{\textsc{i}}$, $\lambda_{\textsc{e}}$, $\kappa_{\textsc{e}}$, and $\mu_{\textsc{t}}$) were generally reliable and accurate. These are the only parameters that should be used to help identify whether the empirical data support the L\'evy walk or the CCRW. No biological interpretation should be based on the probability of starting in the intensive search behavior, $\delta_{\textsc{i}}$. As described by \citet{RefWorks:933}, the estimates from the EM algorithm for this parameter approached either 0 or 1 as $\boldsymbol{\delta}$ will be one of the two unit vectors. Caution should be taken when interpreting the minimum, $a$, and maximum, $b$, step lengths. Even though using the minimum and maximum observed step lengths are the MLEs, and is the suggested method to estimate these values for TLW \citep{RefWorks:940}, some of their estimates were biased. One likely explanation, is that 500 steps was too small a sample to accurately estimate these parameters. The estimates of most parameters of the CCRW suffered when the two search behaviors were not substantially different. 

Because precise methods, such as the likelihood profile, become highly unpractical and computationally demanding when models have more than two or three parameters to be estimated, \citet{RefWorks:934} recommends the use of the quadratic approximation for estimating confidence intervals. Because the CCRW has seven parameters to be estimated, we investigated whether such approximation could be used. The simulation study showed that these approximated confidence intervals were often too narrow and excluded the simulation value. The quadratic approximation can be inaccurate when the parameter estimated is at the boundary of its parameter space \citep{RefWorks:933}. This approximation is symmetric around the MLE, thus might exceed the boundary of parameter space. This occurred for many simulations. For the polar bear data we estimated the confidence intervals using the likelihood profile.

The simulation results showed that our test of absolute fit was adequate, albeit with observed rejection rates that were marginally greater than the expected rate of 0.05. Thus our test had a slightly higher level of type I error than specified by the $\alpha$-level. This problem could be associated with the known negative bias in p-values of G-tests when sample size and expected values are small \citep{RefWorks:225}. We have also explored the use of a number of other tests, such as tests of normality on normal pseudo-residuals \citep[see][for description of normal pseudo-residuals]{RefWorks:933}. None have outperformed the one presented here.

Some sampling procedures, in particular subsampling and the definition of steps by the significant turns, can cause Akaike weights to select L\'evy walk models when CCRWs are simulated \citep{RefWorks:438,RefWorks:928,RefWorks:932}. Although our method is likely to be affected by such procedures, it has features that are known to decrease misidentification errors. In particular, it was shown that including  an approximation of the CCRW and tests for the absolute fit mitigates the risks of such errors \citep{RefWorks:932}. Indeed, through a simulation study, we showed that, while parameter estimates and test of absolute fit were affected by the local turn method with a threshold angle of $10^{\circ}$, our method's capacity to distinguish between the movement patterns of CCRWs and TLWs remained almost unaffected. We have not fully explored the effects of data sampling and handling on the accuracy of our method.  Future work should investigate how sampling procedures impact the capacity of our method to differentiate between the two models.

Overall, our simulation study showed that we can differentiate the L\'evy walk from a strong alternative, such as a CCRW. The Akaike weights could differentiate the L\'evy walk from a CCRW that used a combination of exponential distributions, something that is difficult to accomplish with other methods \citep{RefWorks:440,RefWorks:932,RefWorks:1004}. Other alternative models, including other formulations of the CCRW, could result in movement patterns similar those of a L\'evy walk. In many cases, it might be important to compare the L\'evy walk to a wider range of alternative models. Because Akaike weights are relative measures of fit, it is important to verify that the best model describes the data adequately \citep{RefWorks:930}. Our simulation study demonstrates that our test of absolute fit can identify whether the model describes the data adequately. Finally, our simulation study shows that most parameter estimates are reliable and thus can be used to further investigate whether the data is consistent with the best model. Thus, the simulation study suggests that we could infer support for a model when: 1) it is compared to adequate alternatives, 2) has much higher Akaike weight values than the other models analysed, 3) sufficiently describes the data according to a test of absolute fit, and 4) has parameter estimates consistent with the hypothesis it represents. 

We demonstrated how to interpret the results of our method by applying it to empirical data. Our results suggested that the two bears differed in their movement patterns. For Bear 2, the Akaike weights and parameter estimates suggested that the movement path was better represented by the CCRW. For Bear 1, the Akaike weights suggested that although the CCRW was the best model, the CRW, a one-behaviour null model, might be sufficient to explain the data. These two bears differed in their reproductive status: Bear 1 was accompanied by a yearling at capture while Bear 2 was accompanied by a cub-of-the-year. Females with cubs-of-the-year move smaller distances, avoid adult males to reduce the risk of infanticide, and use lower quality habitat than other bears \citep{RefWorks:1008,RefWorks:1007,Pilfold2014}. Thus, it is possible that females with cubs-of-the-year used different search strategies than other females and this difference could have resulted in the difference observed between the two bears. 

An additional explanation for the difference between these two bears is that the quality of their movement data differ (Fig. \ref{fig:2Bears}). The results for Bear 1 demonstrated that our method can handle large amount of missing data. However, as with most analytical methods, missing data can impact biological interpretation. Specifically, reduced sample size likely hinders our method's ability to differentiate between models and between the two behaviours of the CCRW. In addition, missing locations divides the path into smaller steps, which has the potential to impact model fit. Thus, we advise caution when interpreting results for movement paths with many missing locations. 

The movement path of each bear was significantly different from the best model. This indicates that while better than the other alternatives, the best model is not sufficient to explain polar bear movement. One could easily extend the set of models explored by investigating multiple versions of the HMM. While our choices were made to reduce the number of parameters to be estimated or to ensure that certain characteristics of the CCRW were respected, there are many ways in which the CCRW could be modeled and certain changes could increase its absolute fit. Our CCRW used specific distributions for the frequency of step lengths and turning angles. The choice of such distributions can affect the movement behavior of random walks \citep{RefWorks:987} and other distributions have been used in some multiphasic movement models \citep[e.g., wrapped Cauchy and Weibull distributions, see:][]{RefWorks:443,RefWorks:991}. In addition, by using a simple HMM, we are assuming that the number of steps an animal makes in each behavioral phase follows a geometric distribution. However, the autocorrelation in pseudo-residuals indicates that this assumption might be violated and that a first-order Markov process might be an inaccurate representation of the switching probabilities for polar bears (see Appendix B). One could relax this assumption by using a hidden semi-Markov model \citep{RefWorks:991}. While we explored only one version of a CCRW, our framework allow empiricists to explore a variety of models by simply altering the characteristics of the HMM. 

While exploring a larger variety of CCRWs is likely to increase the absolute fit of our model, it is unlikely that we have sampled the movement paths at the exact scale at which the animals are making their decisions. Sampling scale affects behavioural inference made from movement data \citep[e.g.,][]{RefWorks:974,RefWorks:948,RefWorks:438}. Thus, a lack of strong evidence for the L\'evy walk and CCRW at the scale at which we have sampled our movement paths does not preclude the possibility for such evidence at different scales. Investigating the evidence for these movement models across multiple scales may be useful \citep[e.g.,][]{RefWorks:946,RefWorks:990,Seuront2014}. Finally, it is possible that we are missing important characteristics of polar bear movement. For example, some polar bears move against sea ice drift and ignoring drift can impact interpretation of movement paths \citep{RefWorks:391, RefWorks:400}. Thus, an important extension for polar bears might be the inclusion of drift in the analysis \citep[e.g.,][]{RefWorks:400,Girard2006}.

\subsection{Conclusion}

We have developed likelihood functions for models representing the L\'evy walk and CCRW that make it possible to directly compare the evidence for these two prominent hypotheses. Unlike recently developed methods, our method uses information from both step lengths and turning angles, and incorporates the temporal autocorrelation inherent in the CCRW. Our simulation study showed that our method could differentiate between the two models. By applying our method to the movement path of two polar bears, we showed that our method can give easily interpretable results and handle complex movement paths. The specific model that we used for the CCRW is just one of the many CCRWs that could be created using a HMM. For example, alternate step length and turning angle distributions, such as Weibull and wrapped Cauchy distributions, could be used to create other multi-behavior models with different characteristics \citep[e.g.,][]{RefWorks:991,RefWorks:443}. We hope that application of this method to empirical data will further our understanding of the mechanisms used by animals to find resources. 

\section{Acknowledgments}

We thank Devin Lyons, Craig DeMars, Roland Langrock, and anonymous reviewers for improving the manuscript, Ulrike Schl\"{a}gel, and Alexei Potapov for technical support, and Dennis Andriashek, Nicholas Lunn, and Evan Richardson for fieldwork assistance. We received funding or scholarships from Alberta Innovates-Technology Futures, Aquarium du Qu\'ebec, ArcticNet, Canada Research Chairs program, Canadian Association of Zoos and Aquariums, Canadian Circumpolar Institute, Canadian Wildlife Federation, Environment Canada, Killam trust, Natural Sciences and Engineering Research Council of Canada, Northern Scientific Training Program, Polar Continental Shelf Project, Polar Bears International, Quark Expeditions, Steve and Elaine Antoniuk, University of Alberta, and World Wildlife Fund (Canada).

\bibliographystyle{MEE} 
\bibliography{CBWCRW}

\newpage

\begin{table}\footnotesize
\begin{threeparttable}
	\caption{Likelihood functions and number of parameters to estimates, $k$, of the four models. Table A.1 of Appendix A describes the PDFs, $\phi()$, $\phi_{\textsc{t}}()$, $v()$, $v_{0}()$, and $\psi_{\textsc{t}}()$. Table \ref{tab:AllPar} describes the parameters.}
		\begin{tabular}{l l l}
			\hline
			Model & Likelihood function & $k$ \\
			\hline
			CCRW & 
				$
							
				\begin{smallmatrix}
					(\begin{smallmatrix}
						\delta_{\textsc{i}} & 1-\delta_{\textsc{i}}
					 \end{smallmatrix})\\
					 \mbox{}
				\end{smallmatrix}
							
				\bigl( \begin{smallmatrix}
					\phi(l_1| \lambda_{\textsc{i}},a) \, v_{0}(\theta_1) & 0  \\
					0 & \phi(l_1| \lambda_{\textsc{e}},a) \, v(\theta_1|\kappa_{\textsc{e}})
				\end{smallmatrix} \bigr)
				\;
				\prod^{n}_{i=2}
				\bigl( \begin{smallmatrix}
					\gamma_{\textsc{ii}} & 1 - \gamma_{\textsc{ii}}  \\
					1-\gamma_{\textsc{ee}} & \gamma_{\textsc{ee}}
				\end{smallmatrix} \bigr)
				
				\bigl( \begin{smallmatrix}
					\phi(l_i| \lambda_{\textsc{i}},a) \, v_{0}(\theta_i) & 0  \\
					0 & \phi(l_i| \lambda_{\textsc{e}},a) \, v(\theta_i|\kappa_{\textsc{e}})
				\end{smallmatrix} \bigr)
				\;
				\bigl( \begin{smallmatrix}
					1 \\
					1
				\end{smallmatrix} \bigr)
				$ & 7 \\
			TLW &	$\prod^{n}_{i=1} \; \psi_{\textsc{i}}(l_i|\mu_{\textsc{t}},a,b) \; v_{0}(\theta_i) \; $ & 3 \\
			BW & $\prod^{n}_{i=1}  \; \phi_{\textsc{i}}(l_i|\lambda,a) \; v_{0}(\theta_i) \; $ & 3 \\
			CRW & $\prod^{n}_{i=1} \; \phi_{\textsc{i}}(l_i|\lambda,a) \; v(\theta_i|\kappa) \; $ & 4 \\
			
			\hline
		\end{tabular}
		\label{tab:Models}
		\end{threeparttable}
\end{table}

\clearpage

\newcommand{\tfci}[2]{\begin{tabular}[t]{@{}c@{}} #1 \\[-2ex] \scriptsize (#2) \footnotesize \end{tabular}}
\newcolumntype{P}[1]{>{\raggedright\arraybackslash}p{#1}}
\renewcommand{\tabcolsep}{0.05cm} 

\begin{table}\footnotesize
	\begin{threeparttable}[b]
		\caption{Description and empirical estimates of the model parameters. The parameter estimates and associated confidence intervals (CIs) are presented for each bear.}
			\begin{tabular}{c P{8cm} *{2}{c}}
				\hline
				\tfci{Symbol}{unit} & \multicolumn{1}{c}{Description} & Bear 1 & Bear 2 \\
				\hline
				\tfci{$a$}{$m$} &
				Minimum step length of all four models &
				21 & 2 \\
				\tfci{$b$}{$m$} &
				Maximum step length of the TLW &
				12614 & 11789 \\
				$\delta_{\textsc{i}}$ &
				Probability of starting in the CCRW's intensive search &
				0 & 0 \\
				$\gamma_{\textsc{ii}}$ &
				Probability of remaining in the CCRW's intensive search &
				\tfci{0.48}{0.00-0.74} &
				\tfci{0.83}{0.75-0.89} \\
				$\gamma_{\textsc{ee}}$ &
				Probability of remaining in the CCRW's extensive search &
				\tfci{0.91}{0.83-0.97} &
				\tfci{0.97}{0.95-0.98} \\
				$\kappa$ &
				Size of the directional correlation of the CRW &
				\tfci{1.41}{1.18-1.68} &
				\tfci{1.21}{1.10-1.34} \\
				$\kappa_{\textsc{e}}$ &
				Size of the directional correlation of the CCRW's extensive search &
				\tfci{1.74}{1.41-2.12} &
				\tfci{1.22}{1.09-1.36} \\
				\tfci{$\lambda$}{$m^{-1}$} &
				Rate parameter of the exponential distribution of the BW and CRW &
				\tfci{0.0009}{0.0008-0.0011} &
				\tfci{0.0010}{0.0009-0.0010} \\
				\tfci{$\lambda_{\textsc{i}}$}{$m^{-1}$} &
				Rate parameter of the CCRW's intensive search &
				\tfci{0.0031}{0.0015-0.0059} &
				\tfci{0.0100}{0.0075-0.0131} \\
				\tfci{$\lambda_{\textsc{e}}$}{$m^{-1}$} &
				Rate parameter of the CCRW's extensive search & 
				\tfci{0.0008}{0.0007-0.0010} &
				\tfci{0.0008}{0.0008-0.0009} \\
				$\mu_{\textsc{t}}$ &
				Scale parameter of the truncated Pareto distribution of the TLW &
				\tfci{1.000}{1.000-1.037} & 
				\tfci{1.000}{1.000-1.002} \\
				\hline
			\end{tabular}
	\label{tab:AllPar}
		\end{threeparttable}

\end{table}

\clearpage

\begin{table}\footnotesize
\begin{threeparttable}
	\caption{Relative and absolute fit of the four models on the movement paths of two polar bears.  For each bear, the $\Delta AICc$ and Akaike weight for each model, the p-value for the test of absolute fit of the best model according to AICc, and the number of steps of the movement path are included.}
		\begin{tabular}{*{13}{c}}
			\hline
			Individual & n & \multicolumn{5}{c}{$\Delta AICc$} & 
			\multicolumn{5}{c}{Akaike weight} &
			p-value \\
			& & CCRW & TLW & BW & CRW & & CCRW & TLW & BW & CRW & & Best model\\
			\hline
			Bear 1 & 235 & 0 & 302.4 & 170.3 & 0.4 & & 0.55 & $<0.01$ & $<0.01$ & 0.45 & & $<0.01$ \\
			Bear 2 & 887 & 0 & 1479.7 & 648.6 & 137.2 & & 1.00 & $<0.01$ & $<0.01$ & $<0.01$ & & $<0.01$ \\
			\hline
			\end{tabular}
	\label{tab:EmpModelFitH}
		\end{threeparttable}
\end{table}

\clearpage

\begin{figure}
	\centering
		\includegraphics[width=1.00\textwidth]{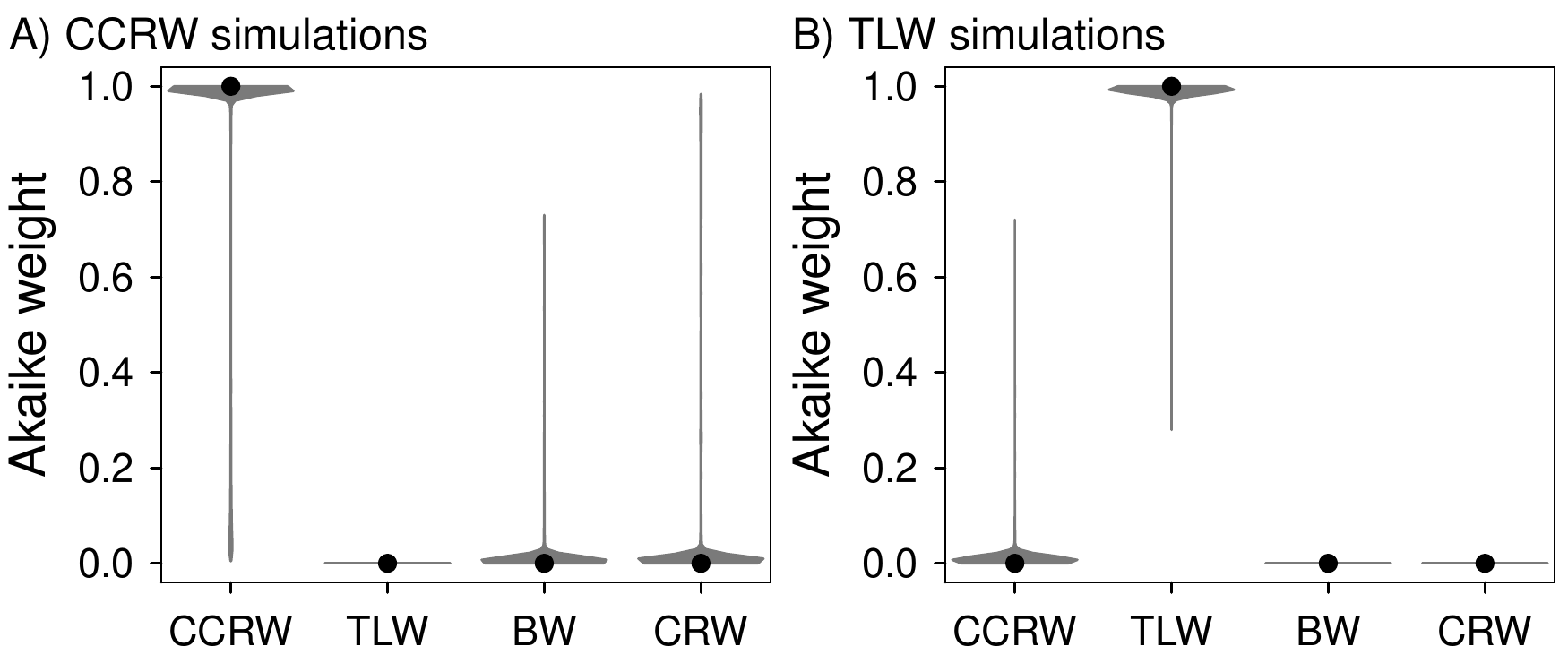}
	\caption{Violin plots of the Akaike weights of each model for all simulated CCRWs and TLWs. High Akaike weight values represent strong support for a model relative to the other models. Violin plots are combinations of kernel density plots (gray polygon) and box plots. Because the range of most model values was orders of magnitude smaller than the y-axis, the box plots are only represented by the $\bullet$ symbols that identify the median. Panels A shows that for simulated CCRWs mostly CCRW had strong support. Panels B shows that for simulated TLWs mostly the TLW had strong support.}
	\label{fig:aw_4M}
\end{figure}

\begin{figure}[ht]
	\centering
		\includegraphics[width=1.00\textwidth]{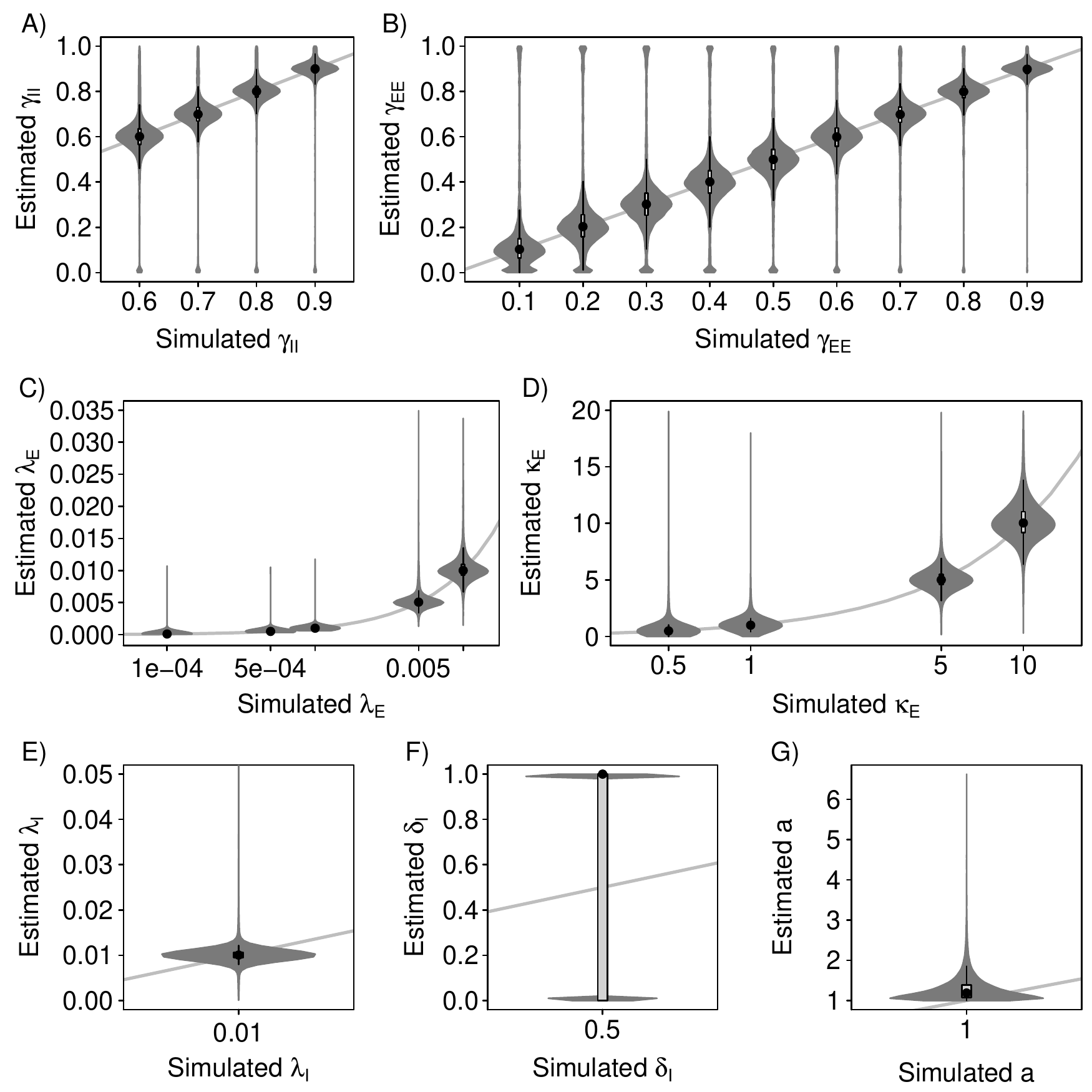}
	\caption{Violin plots of the MLE values for the CCRW simulations. The x- and y-axis represent respectively the values used in the simulations and those recovered by MLE.  The gray line shows their one-to-one relationship. (A) Probability of remaining in intensive search. (B) Probability of remaining in extensive search. (C) Rate parameter of extensive search. (D) Directional correlation of extensive search. (E) Rate parameter of intensive search. (F) Probability of starting in the intensive search. (G) Minimum step length. For visualization, we have cropped out extreme outliers from the plots of $\lambda_{\textsc{i}}$, $\lambda_{\textsc{e}}$, and $\kappa_{\textsc{e}}$, but we removed $< 0.9\%$ of results per parameter value.}
	\label{fig:HMM_MLE}
\end{figure}

\begin{figure}[ht]
	\centering
		\includegraphics[width=1.00\textwidth]{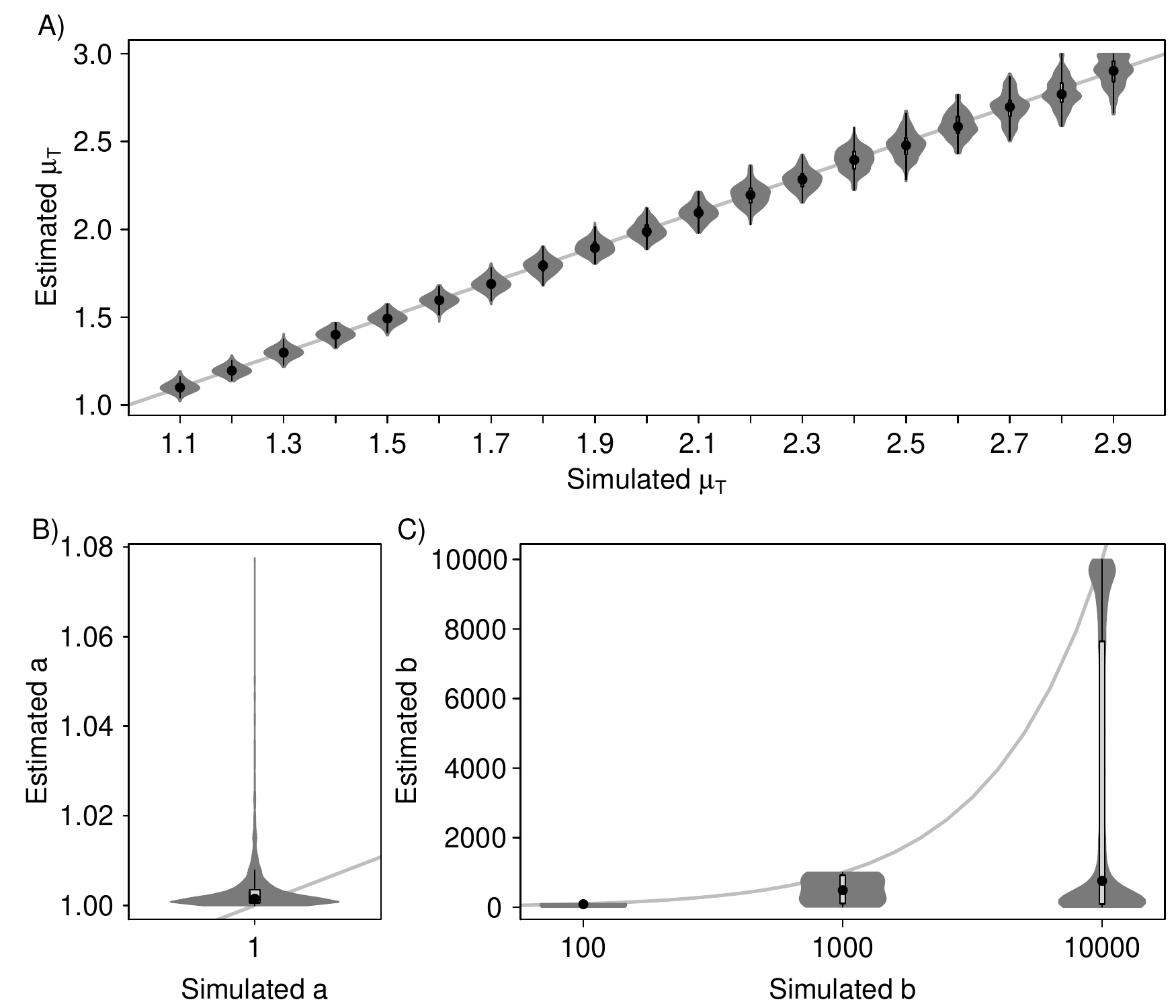}
	\caption{Violin plots of the MLE values of the TLW simulations. The x- and y-axis represent respectively the values used in the simulations and those recovered by MLE. The gray line represents their one-to-one relationship. (A) Scale parameter of the truncated Pareto. (B) Minimum step length. (C) Maximum step length. Estimated values of $\mu_{\textsc{t}}$ are restricted between 1 and 3.}
	\label{fig:TLW_MLE}
\end{figure}

\begin{figure}[ht]
	\centering
		\includegraphics[width=1.00\textwidth]{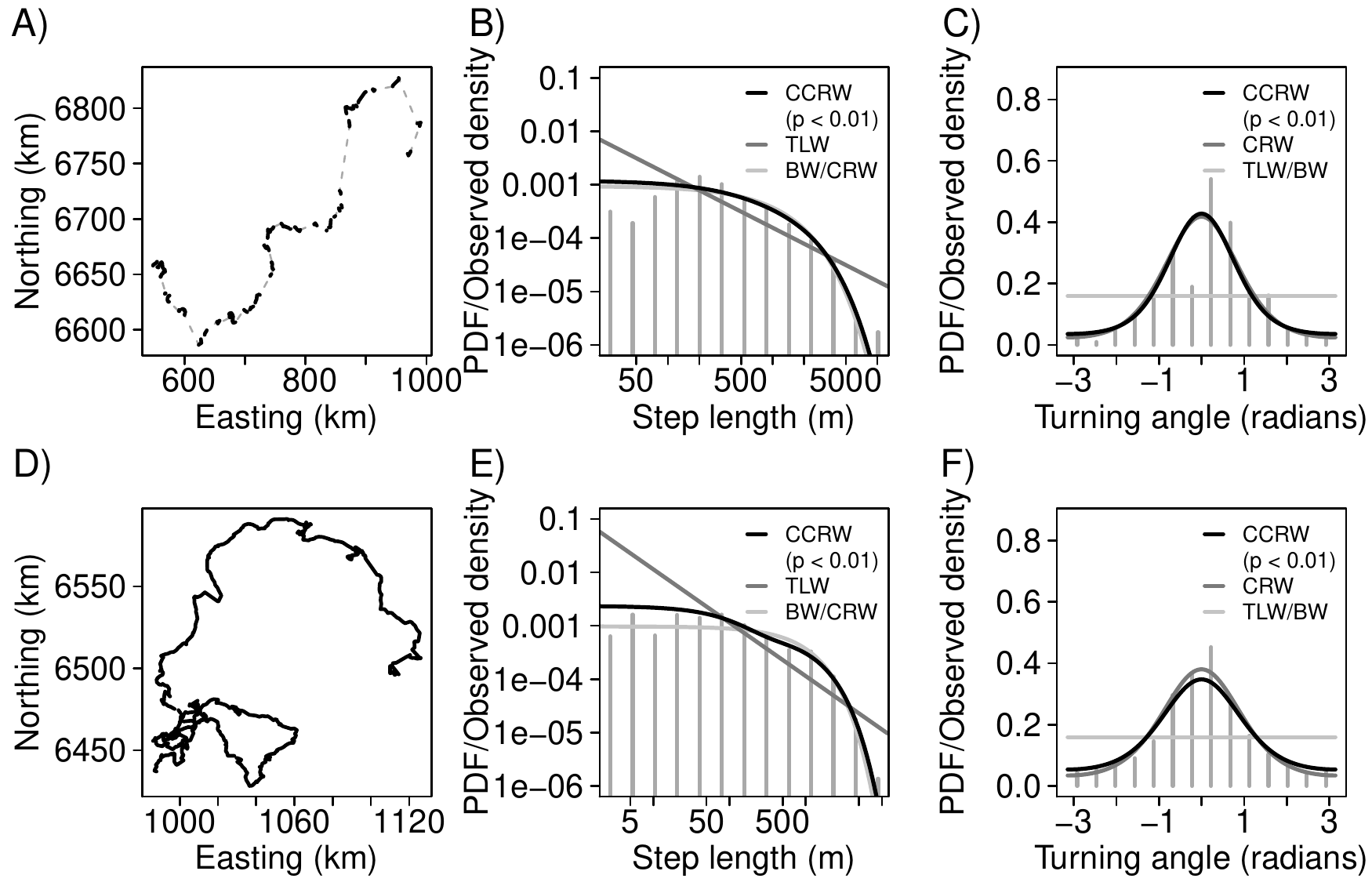}
	\caption{Fit of the models on the movement paths of two polar bears. (A-C) Bear 1. (D-F) Bear 2. (A, D) Movement path, with black lines representing the steps and the dotted line the missing data. (B, E) Step length frequency with the PDF of each model, on log-log axes. (C, F) Turning angle frequency with the PDF of each model. The p-value of the test of absolute fit for the step length and turning angle distributions of the best model are indicated in the legend.}
	\label{fig:2Bears}
\end{figure}

\end{flushleft}
\end{spacing}
\end{document}